\documentclass[aps,pre,twocolumn,showpacs,superscriptaddress,letterpaper]{revtex4}
\usepackage[dvips]{graphicx}
\usepackage{psfrag}
\usepackage{bm}
\usepackage{amsmath}
\usepackage{psfrag}
\usepackage[normalem]{ulem}
\usepackage{pstricks}
\usepackage{xspace}

\newcommand{\ie}{i.e.\@\xspace}
\newcommand{\aka}{a.k.a.\@\xspace}
\newcommand{\eg}{e.g.\@\xspace}

\newcommand{\erf}{\textrm{erf}}
\newcommand{\eq}[1]{Eq.~\eqref{#1}}
\newcommand{\Eq}[1]{Equation~\eqref{#1}}
\newcommand{\eqs}[2]{Eqs.~\eqref{#1} and~\eqref{#2}}
\newcommand{\fig}[1]{Fig.~\ref{#1}}

\newcommand{\figs}[2]{Figs.~\ref{#1} and~\ref{#2}}

\newcommand{\ud}{\mathrm{d}}
\newcommand{\bra}{\left\langle}
\newcommand{\ket}{\right\rangle}
\newcommand{\keff}{k_{\textrm{eff}}}

\newcommand{\imag}{\operatorname{Im}}
\newcommand{\real}{\operatorname{Re}}
\newcommand{\grad}{\vec{\nabla}}
\renewcommand{\div}{\vec{\nabla}\cdot}
\newcommand{\uu}{k_0\ell_{c}}

 \newcommand{\dd}{
      \mathop{}\mathopen{}\mathrm{d}
   }

\newcounter{tempa}
\newcounter{tempb}
\newcounter{tempc}

\newenvironment{diaga}[1]{\psset{unit=1.2mm,fillstyle=solid,fillcolor=white}
   \begin{pspicture}[shift=-2.0](0,-3)(#1,3)}{\end{pspicture}
}
\newenvironment{diagb}[1]{\psset{unit=1.2mm,fillstyle=solid,fillcolor=white}
   \begin{pspicture}[shift=-2.0](0,-3)(#1,6)}{\end{pspicture}
}

\newcommand{\ligne}[2]{\psline(#1,0)(#2,0)}

\newcommand{\particule}[1]{\pscircle(#1,0){1}}

\newcommand{\correldeux}[2]{
   \setcounter{tempa}{#2}
   \addtocounter{tempa}{-#1}
   \divide \value{tempa} by 2
   \setcounter{tempb}{#1}
   \addtocounter{tempb}{#2}
   \divide \value{tempb} by 2
   \psarc[linestyle=dashed](\value{tempb},0){\value{tempa}}{0}{180}
}

\begin{document}

\title{Scattering mean-free path in continuous complex media: beyond the Helmholtz equation}

\author{Ibrahim Baydoun}
\author{Diego Baresch}
\author{Romain Pierrat}
\author{Arnaud Derode}
\affiliation{ESPCI ParisTech, PSL Research University, CNRS, Univ Paris Diderot,
Sorbonne Paris Cit\'{e}, Institut Langevin, 1 rue Jussieu, F-75005, Paris, France}
\email{arnaud.derode@espci.fr}

\date{\today}

\begin{abstract}
   We present theoretical calculations of the ensemble-averaged (\aka effective or coherent)
   wavefield propagating in a heterogeneous medium considered as one realization
   of a random process. In the literature, it is usually assumed that heterogeneity can be accounted
   for by a random scalar function of the space coordinates, termed the potential.
   Physically, this amounts to replacing the constant wavespeed in Helmholtz'
   equation by a space-dependent speed. In the case of acoustic waves, we show that this approach leads to incorrect results for the scattering mean-free path, no matter how weak fluctuations are. The detailed calculation of the coherent wavefield must take into account both a scalar and an operator part in the random potential. When both terms have identical amplitudes, the correct value for the scattering mean-free paths is shown to be more than four times smaller (13/3, precisely) in the low frequency limit, whatever the shape of the correlation function.  Based on the diagrammatic approach of multiple scattering, theoretical results are obtained for the self-energy and mean-free path, within Bourret's and on-shell approximations. They are confirmed by numerical experiments. 
\end{abstract}

\pacs{43.20.+g, 42.25.Dd, 43.35.+d, 46.65.+g} 

\maketitle

\section{Introduction}
% ====================

Whether quantum or classical, electromagnetic or acoustic, wave phenomena share a common
theoretical ground. Hence the universality of fundamental concepts such as coherence, ballistic to diffuse transition, localization, and the observation of related experimental manifestations in all fields of
mesoscopic wave physics ~\cite{SHENG-1995, SEBBAH-2001, SKIPETROV-2003,AKKERMAN-2007}.

In this paper, we are interested in the coherent field i.e., the statistical average of the
wavefield propagating in an inhomogeneous medium whose characteristic are treated as random variables. In mesoscopic physics, determining the coherent field is the very basis of multiple scattering theory. It allows to define a scattering mean-free
path $\ell_s$, which is the key-parameter in any multiple-scattering problem. $\ell_s$ is the typical decay length for the intensity of the coherent wave. In the case of a dilute suspension of discrete scatterers embedded in an homogeneous fluid, $1/\ell_s$ is equal to the scattering cross-section of a single scatterer, multiplied by the number of scatterers per unit volume \cite{FOLDY-1945}. 

Unlike discrete media, what we consider here is an inhomogeneous medium which varies continuously in space. In that case, heterogeneity can be characterized by a random function of the spatial coordinates, called the potential. In this paper, we show that the classical approach to express $\ell_s$ as a function of the fluctuations $\sigma$ and correlation length $\ell_c$ of the potential is incorrect. This is due to an additional term in the acoustic wave equation, which is usually overlooked.  When both terms have identical amplitudes, the correct value for the scattering mean-free paths is four times smaller (13/3, precisely) in the low frequency limit, no matter how weak the fluctuations are and whatever the shape of the correlation function (as long as its second-order moment is finite). As a result, even in the most simple cases (e.g., exponentially-correlated disorder) the scattering mean-free path can be severely underestimated.   

The theoretical framework of the present paper is the diagrammatic approach of multiple scattering~\cite{AKKERMAN-2007, FRISCH-1968, KRT-1989}. It yields an exact
equation for the coherent field known as Dyson's equation, the essential ingredient of which is the self-energy
$\Sigma$. Unfortunately, as often in real life, one has to resort to some degree of approximation to evaluate $\Sigma$
and obtain tractable expressions for the coherent field. The coherent wave has been extensively studied in the literature with various kinds of waves, both theoretically and experimentally ~\cite{ISHIMARU-1982,PAGE-1997,HENYEY-1999,HESPEL-2001,LINTON-2005,DERODE-2006,PAGE-2011,LUPPE-2012}.

From a theoretical point of view, in the case of continuous heterogeneous media the starting point is usually a wave equation in which disorder is introduced by a space-dependent wave velocity $c(\vec{r})$:
\begin{equation} \label{Eq_Ondes_rt}
   \Delta \phi - \frac{1}{c^2}\frac{\partial^2 \phi}{\partial t^{2}}=s(\vec{r},t).
\end{equation}
Assuming linearity and time-invariance, in order to determine the wavefield $\phi(\vec{r},t)$ generated by any distribution of sources $s(\vec{r},t)$ it suffices to know the Green's function $g(\vec{r},\vec{r_s},t)$ \ie, the solution of \eq{Eq_Ondes_rt} when the source term is $s(\vec{r},t)=\delta(\vec{r}-\vec{r_s})\delta(t)$, with appropriate boundary conditions.

An essential point is that in \eq{Eq_Ondes_rt} heterogeneity is fully characterized by a random \textit{scalar} $c(\vec{r})$. The average Green's function $\langle g\rangle$, hence the coherent field $\langle\phi\rangle$, can be calculated, provided the statistical properties of $c(\vec{r})$, particularly its correlation function, are known.

Here we are interested in a very simple case in which the usual approach fails. We consider an acoustic wave propagating
in a lossless heterogenous fluid. It is well known that the wave equation for the acoustic pressure does not take the
form of \eq{Eq_Ondes_rt}: it entails an additional term with a random \textit{operator}, instead of a simple
scalar~\cite{CHERNOV-1960,JENSEN-2011,ROSS-1986}. The operator term is usually neglected when dealing with multiple
scattering of waves. This implies an important error in the calculation of the mean-free path, especially at low
frequencies. To our knowledge, this point has been overlooked so far. Let us mention however Ref.~\cite{PAVLOV-1992}, in
the context of acoustic Cerenkov radiation by a moving source in a turbulent medium. Turner \emph{et al.}~\cite{TURNER-2001} derived a Dyson equation in the case of isotropic solids with weak fluctuations of density and Lam\'{e} coefficients; yet the liquid limit (no shear modulus) of this model does not exactly yield the correct mean-free path for a fluid, as will be discussed later. 

The paper is organised as follows. In the next section we briefly recall the basics of the diagrammatic approach of multiple scattering, applied to the standard wave equation, and show why this is inappropriate in the case of acoustic waves. Section \ref{CalculsOperateur} gives a complete solution of the wave equation for the average acoustic pressure. Under the Bourret approximation, we show that the self-energy $\Sigma$ exhibits three additional terms compared to the standard scalar case. Therefore section \ref{CalculsOperateur}, particularly \eq{Sigma_total_x}, is the core of the paper. In the final section, we discuss the importance of these additional terms in the simple situation of an exponentially-correlated disorder. Numerical experiments based on averages of numerous simulations of the complete wave equation are found in very good agreement with the theoretical results. Supplementary calculations and numerical details are presented in the Appendix.

\section{The usual way and its limitations}
% ====================================

Let us consider first a homogenous and lossless medium, with a constant sound speed $c_{0}$. The corresponding Green's function and its Fourier transforms will be denoted $g_0$, $G_0$ and $\tilde{G}_0$ respectively. In the monochromatic regime, $G_0$ is the solution of Helmholtz' equation:
\begin{equation} \label{Helm_homo}
   \Delta G_0 + \dfrac{\omega^2}{c_0^2} G_0=\delta(\vec{r}-\vec{r_s}).
\end{equation}
$\omega$ is the angular frequency, and $\omega/c_{0}=k_{0}$ is a reference wavenumber. With the condition of causality, in unbounded 3-D space we have: 
\begin{align} \label{G0}
   g_{0}(\vec{r},\vec{r}_{s},t)&=-\dfrac{1}{4\pi |\vec{r}-\vec{r}_{s}|} \delta(t-|\vec{r}-\vec{r}_{s}|/c_0),
      \\
   G_{0}(\vec{r},\vec{r}_{s},\omega) &=-\dfrac{1}{4\pi |\vec{r}-\vec{r}_{s}|} e^{jk_{0}|\vec{r}-\vec{r}_{s}|},
   \\
   \tilde{G}_{0}(\vec{k}, \omega) &=\dfrac{1}{k_0^2-k^2},
\end{align}
with $\vec{k}$ the dual variable for $\vec{r}-\vec{r}_{s}$ and $j=\sqrt{-1}$. The tilde $\tilde{\cdot}$ denotes a spatial Fourier transform. In the sequel, the analysis is performed in the frequency domain and the $\omega$-dependence is dropped for brevity. 

%========================
\subsection{Scalar potential}
%=========================
Assuming that
a heterogeneous medium can be simply characterized by a space-dependence of the wave speed amounts to replacing $c_{0}$
by $c(\vec{r})$ in the wave equation. Then it is convenient to define the scalar potential $\alpha$:
\begin{equation} \label{def_alpha}
   \alpha(\vec{r})=1-c_{0}^{2}/c^{2}(\vec{r}).
\end{equation}
 The monochromatic Green's function $G(\vec{r},\vec{r}_{s})$ in the heterogeneous medium is such that
\begin{equation} \label{Helm_hetero}
   \Delta G + k_{0}^{2} G =k_{0}^{2}\alpha G+\delta(\vec{r}-\vec{r}_{s}).
\end{equation}

Note that in \eq{Helm_hetero} $c_{0}=k_{0}\omega$ is a reference speed which could be chosen arbitrarily. It is often convenient to set 
\begin{equation} \label{c_0}
   \frac{1}{c_{0}^2}=\bra \frac{1}{c^{2}(\vec{r})} \ket,
\end{equation}
so that $\langle \alpha \rangle=0$; the brackets denote an ensemble average.

For media such that the typical speed fluctuation $\delta c$ is much smaller than the average speed $\langle c\rangle$,
\eq{c_0} amounts to choosing $c_{0}=\langle c\rangle$, hence the reference speed is actually the average sound speed,
but this is not true in the general case.

The potential $\alpha$ fully characterizes the heterogeneity, in that it measures the gap between the reference and the actual wave speed, at any point in space. The term \emph{potential} comes from quantum
mechanics, where the relevant wavefield is the complex amplitude of probability
and obeys Schr\"{o}dinger equation, in which case the heterogeneity of the
medium is an actual potential energy~\cite{AKKERMAN-2007}. Here, $\alpha$ is simply a dimensionless
function of the space coordinate $\vec{r}$.

\eq{Helm_hetero} is
similar to \eq{Helm_homo}, with a source term
$\alpha(\vec{r})G(\vec{r},\vec{r}_{s})+\delta(\vec{r}-\vec{r}_{s})$ that entails
the Green's function itself. Hence $G$ can be expressed implicitly in a
recursive manner (Lippmann-Schwinger form) as:
\begin{equation} \label{G_recur}
   G(\vec{r},\vec{r}_{s})=G_{0}(\vec{r},\vec{r}_{s})+k_{0}^{2}\int G_{0}(\vec{r},\vec{r}_{1})
   \alpha(\vec{r}_{1})G(\vec{r}_{1},\vec{r}_{s})\ud^{3}{r}_{1}.
\end{equation}
For an arbitrary source distribution, the resulting wavefield at $\vec{r}$ would be
\begin{equation} \label{phi=G_conv_s}
\phi(\vec{r})=\int G(\vec{r},\vec{r}_{s})
s(\vec{r}_s)d^{3}{r}_{s}.
\end{equation}

Substituting $G$ under the integral by the right-hand side of \eq{G_recur} and
reiterating the process yields an exact expression for $G$ as an infinite
sum of multiple integrals, known as Born's expansion:
\begin{multline} \label{Devpt_Born}
   G(\vec{r},\vec{r}_{s})=G_{0}(\vec{r},\vec{r}_{s})+k_{0}^{2}\int G_{0}(\vec{r},\vec{r}_{1})
      \alpha(\vec{r}_{1})G_{0}(\vec{r}_{1},\vec{r}_{s})\ud^{3}{r}_{1}
\\
   +k_{0}^{4}\iint G_{0}(\vec{r},\vec{r}_{1}) \alpha(\vec{r}_{1})G_{0}(\vec{r}_{1},\vec{r}_{2})
      \alpha(\vec{r}_{2})G_{0}(\vec{r}_{2},\vec{r}_{s})\ud^{3}{r}_{1}\ud^{3}{r}_{2}
\\
   +\ldots
\end{multline}

The single-scattering approximation (which is commonly made in imaging of weakly heterogeneous media) consists in
neglecting all terms beyond the first integral on the right-hand side of \eq{Devpt_Born}. In that case the Green's
function $G$ can be easily computed for any function $\alpha$, and the inverse problem (\ie reconstructing $\alpha$ from
$G$) may be solved. Naturally this approach completely fails as soon as multiple scattering is not negligible.

In this paper, we consider multiple scattering of waves but we do not aim at solving \eq{G_recur}. Considering
$\alpha$ as a random variable with known statistical parameters, we are interested in the statistical average $\langle
\phi\rangle$ of the wavefield, which amounts to determine $\langle
G\rangle$. We will refer to it as the \emph{coherent} field. This
approach implies that we consider a given medium as one particular realization, among the infinity of possible outcomes,
of the same random process. From a physical point of view, what an experimentalist would measure with a source distribution $s(\vec{r}_s)$ and a point receiver at $\vec{r}$ is $\phi$, not $\langle \phi \rangle$. But $\phi$ can be formally written as $\langle \phi \rangle + \delta
\phi $ with $\langle\delta\phi\rangle=0$ (\ie a mean term plus statistical fluctuations, changing from one realisation to an
other). How $\langle \phi \rangle$ can be estimated in an actual experiment, and how robust the estimation is, is not our
concern in the present paper. We focus on theoretical calculations for $\langle \phi\rangle$, derived from the statistical
properties of $\alpha$. By taking the average of the Born expansion, this will obviously require to know the statistical
moments of $\alpha$ with any order $n$ (\ie quantities such as
$\langle\alpha(\vec{r}_{1})\alpha(\vec{r}_{2})\ldots\alpha(\vec{r}_{n})\rangle$).

The diagrammatic theory of multiple scattering shows that $\langle G\rangle$ obeys an exact equation,
known as Dyson's equation~\cite{FRISCH-1968,KRT-1989}. The basic ingredient in Dyson's equation is a quantity $\Sigma$ referred to as the
self-energy or the mass operator in the literature. $\Sigma$ can be fully determined by the statistical properties of
$\alpha$, and takes into account all orders of multiple scattering. In a nutshell, the basic idea is to rewrite the
statistical average of \eq{Devpt_Born} as an implicit, recursive expression for $\langle G\rangle$. The resulting
Dyson's equation reads:
\begin{multline} \label{Dyson_general}
   \langle G(\vec{r},\vec{r}_{s})\rangle = G_{0}(\vec{r},\vec{r}_{s})
\\
   +\iint G_{0}(\vec{r},\vec{r}_{1}) \Sigma(\vec{r}_{1},\vec{r}_{2})
      \langle G(\vec{r}_{2},\vec{r}_{s})\rangle \ud^{3}{r}_{1}\ud^{3}{r}_{2}.
\end{multline}
Assuming that the medium is statistically homogeneous (\ie,
 its statistical parameters are invariant under translation)
$\Sigma(\vec{r}_{1},\vec{r}_{2})$ only depends on $\vec{x}=\vec{r}_{1}-\vec{r}_{2}$, and since so does $G_{0}$,
\eq{Dyson_general} is a double convolution product on the variable $\vec{x}$. Hence it can be simply solved by a
spatial Fourier transform, which yields
\begin{equation} \label{Gmoy_general}
   \langle \tilde G(\vec{k})\rangle = \frac{1}{k_{0}^{2}-k^{2}-\tilde{\Sigma}(\vec{k})},
\end{equation}
where $\vec{k}$ is the dual variable for $\vec{x}$.  Assuming further that the medium is statistically isotropic (i.e.,
its statistical parameters are also invariant under rotation), both $\tilde \Sigma$ and $\tilde G$ only depend on
$k=|\vec{k}|$.

The key issue is naturally to determine $\Sigma$. Mathematically, $\Sigma$ can be written as a pertubative development
in $1/(k_{0}\ell_{s})$, an infinite series of integrals involving statistical moments of $\alpha$ at all orders,
which can be represented by the following diagrams

\begin{equation}  \label{Diagrammes}
   \Sigma =
      \begin{diaga}{2}
         \particule{1}
      \end{diaga}+
      \begin{diaga}{8}
         \correldeux{1}{7}
         \ligne{1}{7}
         \particule{1}
         \particule{7}
      \end{diaga}+
      \begin{diagb}{14}
         \correldeux{1}{13}
         \ligne{1}{7}
         \ligne{7}{13}
         \particule{1}
         \particule{7}
         \particule{13}
      \end{diagb}+\ldots
\end{equation}

Following the usual conventions, a continuous line joining two points represents the free-space Green's function $G_{0}$ between
these points; a dashed line linking $n$ points is the $n$-order moment of $\alpha$ multiplied by $k_{0}^{2n}$. The inner
points of a diagram are dummy variables. For the establishment of \Eq{Diagrammes}, see for instance ~\cite {FRISCH-1968, KRT-1989}.

The Bourret approximation (\aka first-order smoothing approximation) only keeps
the first two terms in the perturbative development of $\Sigma$. This yields a simple analytical expression for $\Sigma$
as a function of the first and second-order moments (\ie the mean $\langle \alpha \rangle$ and correlation function $C_{\alpha
\alpha}(\vec{r}_{1}-\vec{r}_{2})=\langle \alpha(\vec{r}_{1})\alpha(\vec{r}_{2}) \rangle$). Note that the Bourret approximation
does not imply at all that multiple scattering terms are neglected beyond second-order scattering, but rather than all
multiple scattering events are assumed to be similar to a succession of uncorrelated single or double-scattering
sequences. 

Under the Bourret approximation, and having chosen $c_{0}$ such that $\langle
\alpha \rangle=0$, the expression for the self-energy is ~\cite{FRISCH-1968, KRT-1989}:
\begin{equation} \label {Sigma_Bourret}
   \Sigma(\vec{x})=k_{0}^{4} G_{0}(\vec{x}) C_{\alpha \alpha}(\vec{x}).
\end{equation}

The last step is to determine the average Green's function from \eq{Gmoy_general}. In the most general case, it is an
arbitrary function of $\vec{k}$. To determine the average Green's function $\langle G(\vec{x})\rangle$ in real space,
one has to perform an inverse Fourier transform. This is not always possible analytically and does not always lead to a
simple effective medium; $\Sigma$ is said to be non-local. We will not discuss these issues in the present paper.
Instead, we make a further approximation referred to as the on-shell approximation. It usually requires
$\tilde{\Sigma} (\vec{k})$ to be sufficiently smaller than $k_{0}^{2}$. Indeed, if $\tilde{\Sigma}(\vec{k})$ is weak
enough, we can reasonably assume that the effect it will have on the homogeneous wavenumber $k_{0}$ is small, so that
when performing the inverse three-dimensional Fourier transform, the volume that essentially contributes to $\langle
G(\vec{r}-\vec{r}_{s})\rangle$ is the vicinity of the shell defined by $|k|=k_{0}$. In other words, this amounts to
performing a zero-order development of the self-energy around $k_{0}$, hence replacing $\tilde{\Sigma}(\vec{k})$ by
$\tilde{\Sigma}(\vec{k}_{0})$ in \eq{Gmoy_general}. In that case the expression of the average Green's function in
real space is straightforward:
\begin{equation} \label {Gmoy_onshell}
   \langle G(\vec{r}-\vec{r}_{s})\rangle=-\dfrac{1}{4\pi |\vec{r}-\vec{r}_{s}|} e^{j\keff|\vec{r}-\vec{r}_{s}|}.
\end{equation}
This means that the average Green's function is that of a fictitious homogeneous absorbing medium with a complex-valued wavenumber
$\keff$ such that
\begin{equation} \label{k_eff}
   \keff=\sqrt{k_{0}^{2}-\tilde{\Sigma}(\vec{k}_{0})}.
\end{equation}

\Eq{k_eff} is a dispersion relation from which phase and group velocities for the coherent field can be determined. Most importantly,
the intensity of the coherent wave is found to decay exponentially with the distance (Beer-Lambert's law). Since there is no absorption, the losses are  entirely due to scattering and the scattering mean-free path is $\ell_{s}=1/[2\imag(\keff)]$.

$\ell_{s}$ is an
essential parameter in all multiple scattering problems. Particularly, the range of validity of the Bourret
approximation can be shown to be $k_{0}\ell_{s}\gg1$ ~\cite{KRT-1989}. It should also be mentioned that, as a refinement of the on-shell
approximation, $\keff$ can be determined more accurately with an iterative algorithm using $k_0$ as a first guess.

So, within the Bourret approximation, as long as the correlation function for the potential $\alpha$ is known, the
effective wave speed and scattering mean-free path can be determined quite easily and sometimes analytically.

A typical
example is that of a disordered random medium with an exponentially decaying correlation function $C_{\alpha
\alpha}(\vec{x})=\sigma^{2}_{\alpha}\exp(-x/\ell_{c})$. $\sigma^{2}_{\alpha}$ denotes the variance of $\alpha$ and $\ell_{c}$ its
correlation length. In that case, under the Bourret approximation the self-energy is:
\begin{equation} \label {Sigma_cas_expo}
   \tilde{\Sigma}(\vec{k})=-\frac{\sigma^2_{\alpha}k_{0}^{4}}{k^{2}+(jk_{0}-1/\ell_{c})^{2}}.
\end{equation}
The on-shell approximation yields 
\begin{equation}
	\keff=k_0\sqrt{1+\dfrac{(\sigma_\alpha k_0\ell_c)^2}{1-2jk_0\ell_c}},
\end{equation}
whose imaginary part determines the scattering mean-free path. Furthermore, if $\Sigma(k_0)\ll k_0^2$, a first-order development of the square-root gives a simple analytic expression for the scattering mean-free path $\ell_s$ as a function of frequency, correlation length and variance of the potential: 
\begin{equation} \label {ls_cas_expo}
   \ell_s=\frac{1}{\sigma_\alpha^2 k_0}\frac{1+4k_0^2\ell_c^2}{2k_0^3\ell_c^3},
\end{equation}
allowing us to work with practical dimensionless quantities and express $k_0\ell_s\sigma_\alpha^2$ as a function of $k_0\ell_c$.
Beyond this simple example, whatever the shape of the correlation function and whatever the nature of the wave (acoustic, electromagnetic, \ldots), the same formalism will hold as soon as we deal with a  wavefield propagating in a heterogeneous medium in which heterogeneity is fully described by a scalar function such as $\alpha$. Such is the case when the local wavespeed
$c(\vec{r})$ suffices to capture the heterogeneity, which is usually assumed as a starting point in multiple scattering
theories. However this description of heterogeneity may sometimes be completely misleading, even in very simple
situations.

%========================
\subsection{Operator potential}
%=========================
Let us consider the case of acoustic waves in a lossless fluid. The medium is characterized by its mass
density $\rho$ and compressibility $\chi$ at rest. With no sources, the linearized equations of elastodynamics read
\begin{align} \label{Eq_base1}
   \rho\frac{\partial \vec{v}}{\partial t}= &\, -\grad p,
\\ \label{Eq_base2}
   \div\vec{u}= &\, -\chi p.
\end{align}
$\vec{u}(\vec{r},t)$ is the displacement undergone by the particle initially at rest at point $\vec{r}$,
$\vec{v}=\partial \vec{u}/\partial t$ is the particle velocity, and $p(\vec{r},t)$ is the acoustic pressure. $\vec{u}$,
$\vec{v}$ and $p$ are first-order infinitesimal quantities. To establish \eqs{Eq_base1}{Eq_base2} all second-order
non-linear terms have been neglected whatever their physical origin (convective or thermodynamic). From a physical point
of view, \eqs{Eq_base1}{Eq_base2} are an expression of Newton's second law and Hooke's law (the relative dilation
$\div\vec{u}$ undergone by an infinitesimal volume of fluid is proportional and opposed to the acoustic pressure). If
neither $\rho$ nor $\chi$ depend on space coordinate $\vec{r}$, then
\eqs{Eq_base1}{Eq_base2} lead to the usual wave equation with a constant sound velocity
$c_{0}=1/\sqrt{\rho\chi}$, which applies to all quantities describing the sound wave (acoustic pressure,
velocity, displacement, dilation, etc\ldots). 

In a heterogeneous fluid, the local sound velocity naturally depends on the space coordinate $\vec{r}$ as
$c(\vec{r})=1/\sqrt{\rho(\vec{r})\chi(\vec{r})}$. It is therefore tempting to replace $c_0$ by $c(\vec{r})$ in Helmholtz' equation for a homogeneous fluid (\eq{Helm_homo}), but this is not always correct.

Combining \eqs{Eq_base1}{Eq_base2} yields the following equations for the acoustic pressure and velocity:
\begin{align} \label{eq_p_hetero}
   &\Delta p - \dfrac{1}{c^{2}}
   \dfrac{\partial ^{2}p}{\partial t^{2}}-\frac{\grad\rho\cdot\grad p}{\rho} =0,
   \\
   \label{eq_v_hetero}
   	&\vec{\Delta} \vec{v} - \dfrac{1}{c^2}
   	\dfrac{\partial^{2}\vec{v}}{\partial t^2}
   	-\frac{\grad \chi \cdot \grad \cdot \vec{v}}{\chi} =0.
\end{align}

If $\rho$ does not vary in space, \eq{eq_p_hetero} yields the usual wave equation for the acoustic pressure $p$, with a space-dependent velocity $c$. And if $\chi$ does not vary in space, \eq{eq_v_hetero} yields the usual wave equation for the velocity $\vec{v}$. But in the general case where \emph{both}
$\rho$ and $\chi$ vary with $\vec{r}$, none of the acoustic variables satisfy the usual wave equation ~\cite{CHERNOV-1960, JENSEN-2011}. However the resulting equation for the acoustic pressure (\eq{eq_p_hetero}) can be Fourier-transformed over time, then rearranged in order to define a potential, as we did before. We
obtain an equation similar to Helmholtz':
\begin{equation} \label{Helm_hetero_2}
   \Delta P + k_{0}^{2}P =k_{0}^{2} \gamma P.
\end{equation}
The potential $\gamma$ is such that:
\begin{equation} \label{def_gamma}
   \gamma(\vec{r}) = \alpha(\vec{r}) +\frac{1}{k_{0}^{2}} \grad\beta(\vec{r})\cdot\grad.
\end{equation}
$\alpha$ is defined in \eq{def_alpha} and
\begin{equation} \label{def_beta}
   \beta(\vec{r}) = \ln\left[\frac{\rho}{\rho_0}\right].
\end{equation}
$\rho_0$ is an arbitrary constant with the dimensions of a mass density.

The first term in the expression of $\gamma$ is the usual potential $\alpha$, related to the spatial fluctuations of sound
speed. In addition, there is an other term, which unlike $\alpha$ is not a simple scalar but an \emph{operator} acting on
the field it is applied to. 
It should be noticed that the same problem arises for different kinds
of waves \eg, electromagnetic waves propagating in a heterogeneous medium showing fluctuations of both relative permeability
$\mu(\vec{r},\omega)$ and permittivity $\epsilon(\vec{r},\omega)$.
In that case, Maxwell's equations yield the following wave equation for the monochromatic
electric field $\vec{E}(\vec{r},\omega)$, with a potential that contains an operator part~\cite{BORN&WOLF-1999}:
\begin{equation} \label{Helm_EM}
   \grad\times\grad\times\vec{E} - k_{0}^{2}\epsilon\mu\vec{E} -\grad\left[\ln(\mu)\right]\times\grad\times\vec{E} =\vec{0}.
\end{equation}

Note that the real issue is not to determine when fluctuations in permeability (or density, in the acoustic case) can be neglected compared to fluctuations in permittivity (resp.~compressibility): both kinds of fluctuations, whether separated or combined,  will result in a space-dependent wave speed $c(\vec{r})$.  The question should rather be set in terms of comparing the operator part and the scalar part in the potential describing the heterogeneity.

Whatever the physical nature of the wave, the applicability of the diagrammatic techniques when the relevant potential $\gamma$ has both a scalar part and an operator part as well as the impact of the operator part on the final result are not trivial. In the next section, we deal with this problem and obtain the average Green's function, in the case of acoustic waves.

\section{Self-energy calculation}\label{CalculsOperateur}
% =======================

In this section, the acoustic pressure $P$ is chosen as the relevant variable for the wavefield. The issue is to determine the average
Green's function of \eq{Helm_hetero_2}. The reference velocity is still chosen according to \eq{c_0}. Since
the medium is assumed to be statistically invariant under translation $\langle \beta(\vec{r})\rangle$ does not depend on
the space coordinate $\vec{r}$. Hence, despite the presence of $\beta$, the average value of the potential $\gamma$ will
still be zero, since
\begin{equation}
   \langle \grad\beta \rangle= \grad\langle \beta \rangle=\vec{0}.
\end{equation}
As usual, the Green's function for \eq{Helm_hetero_2} can be written as a Lippmann-Schwinger equation
\begin{equation} \label{G_recur_2}
   G(\vec{r},\vec{r}_{s})=G_{0}(\vec{r},\vec{r}_{s})+k_{0}^{2}\int G_{0}(\vec{r},\vec{r}_{1})
   \gamma(\vec{r}_{1})G(\vec{r}_{1},\vec{r}_{s})\ud^{3}{r}_{1}.
\end{equation}
$\gamma$ is not a simple scalar function, which precludes the usual definition of its
autocorrelation function and that of the self-energy. To overcome this difficulty, we start by introducing a
two-variable potential $V$ such that
\begin{align} \label{def_V}
   \nonumber
   V(\vec{r}_{1},\vec{r}_{2})&=\gamma(\vec{r}_{1})\delta(\vec{r}_{1}-\vec{r}_{2})
   \\&
  = \alpha(\vec{r}_{1})\delta(\vec{r}_{1}-\vec{r}_{2})
         +\frac{1}{k_{0}^{2}}\grad_{\vec{r}_{1}}\beta(\vec{r}_{1})\cdot\grad_{\vec{r}_{1}}\delta(\vec{r}_{1}-\vec{r}_{2}).
\end{align}

\eq{G_recur_2} becomes
\begin{multline} \label{G_recur_3}
   G(\vec{r},\vec{r}_{s})=G_{0}(\vec{r},\vec{r}_{s})
\\
   +k_{0}^{2}\iint G_{0}(\vec{r},\vec{r}_{1})
      V(\vec{r}_{1},\vec{r}_{2})G(\vec{r}_{2},\vec{r}_{s})\ud^{3}{r}_{1}\ud^{3}{r}_{2}.
\end{multline}
The next steps are as usual to develop \eq{G_recur_3} into a Born expansion by iteration, then to take its
statistical average and write it under a recursive form (Dyson's equation). Under the Bourret approximation, only the
first two terms in the self-energy are kept. The first one vanishes since $c_{0}$ is set so that $\left\langle \gamma\right\rangle=0 $. The second
term reads:
\begin{equation} \label {Sigma_Bourret_beta}
   \Sigma(\vec{r}_{a},\vec{r}_{b})=k_{0}^{4}\iint \langle V(\vec{r}_{a},\vec{r}_{1}) G_{0}(\vec{r}_{1},\vec{r}_{2})
   V(\vec{r}_{2},\vec{r}_{b})\rangle \ud^{3}{r}_{1}\ud^{3}{r}_{2}.
\end{equation}

As a consequence, the self-energy \eq{Sigma_Bourret_beta} gives rise to four terms, involving the
following dimensionless correlation functions and their derivatives:
\begin{equation} \label {4 fcts de corrélation}
   \begin{split}
      C_{\alpha \alpha}(\vec{r}_{1},\vec{r}_{2})= &\, \langle \alpha(\vec{r}_{1})\alpha(\vec{r}_{2}) \rangle,
   \\
      C_{\beta \beta}(\vec{r}_{1},\vec{r}_{2})= &\, \langle \beta(\vec{r}_{1})\beta(\vec{r}_{2}) \rangle,
   \\
      C_{\alpha \beta}(\vec{r}_{1},\vec{r}_{2})= &\, \langle \alpha(\vec{r}_{1})\beta(\vec{r}_{2}) \rangle,
   \\
      C_{\beta \alpha}(\vec{r}_{1},\vec{r}_{2})= &\, \langle \beta(\vec{r}_{1})\alpha(\vec{r}_{2}) \rangle.
   \end{split}
\end{equation}
We assume that the random processes $\alpha$ and $\beta$ are \emph{jointly} stationary and invariant under rotation. Then
all correlation functions only depend on $x=|\vec{r}_{1}-\vec{r}_{2}|$ and $C_{\alpha \beta}(x)=C_{\beta \alpha}(x)$. Only three
correlation functions suffice to characterize the disorder. They can be rewritten as
\begin{equation} \label {3 fcts de corrélation}
   \begin{split}
      C_{\alpha \alpha}(x)= &\, \sigma_{\alpha}^{2}c_{\alpha\alpha}(x),
   \\
      C_{\beta \beta}(x)= &\, \sigma_{\beta}^{2}c_{\beta\beta}(x),
   \\
      C_{\alpha \beta}(x)=C_{\beta \alpha}(x)= &\, \sigma_{\alpha}\sigma_{\beta}c_{\alpha\beta}(x).
   \end{split}
\end{equation}
$\sigma_{\alpha}^{2}$, $\sigma_{\beta}^{2}$ are the variances of $\alpha$ and $\beta$ respectively. $c_{\alpha\alpha}(x)$ is the
correlation coefficient between $\alpha(\vec{r})$ and $\alpha(\vec{r}+\vec{x})$. $c_{\beta\beta}(x)$ is the correlation
coefficient between $\beta(\vec{r})$ and $\beta(\vec{r}+\vec{x})$. $c_{\alpha\beta}(x)$ is the correlation coefficient between
$\alpha(\vec{r})$ and $\beta(\vec{r}+\vec{x})$.  Replacing $V$ in \eq{Sigma_Bourret_beta} by \eq{def_V}, we can write the
self-energy $\Sigma$ as a sum of four contributions:
\begin{equation} 
   \Sigma=\Sigma_{\alpha \alpha}+\Sigma_{\beta \alpha}+\Sigma_{\alpha \beta}+\Sigma_{\beta \beta}.
	\label{eq:Self}
\end{equation}

\subsection{First term}\label{contrib_mumu}
% ---------------------

The first term is:
\begin{align} \label {Sigma_mumu}
   \Sigma_{\alpha\alpha}(\vec{r}_{a}-\vec{r}_{b})= &\, k_{0}^{4} \langle \alpha(\vec{r}_{a}) G_{0}(\vec{r}_{a}-\vec{r}_{b}) \alpha(\vec{r}_{b})\rangle
\\
   = &\, k_{0}^{4}G_{0}(\vec{r}_{a}-\vec{r}_{b})C_{\alpha \alpha}(\vec{r}_{a}-\vec{r}_{b}).
\end{align}
With $\vec{x}=\vec{r}_{a}-\vec{r}_{b}$ we find the usual contribution to the self-energy \eq{Sigma_Bourret}, involving only the scalar potential $\alpha$.

The
three additional terms are more complicated, since they entail combinations of $\alpha$ and $\beta$ as well as spatial
derivatives.

\subsection{Second term}\label{contrib_numu}
% ----------------------

The second term mixes contributions from $\beta$ and $\alpha$:
\begin{multline}
   \Sigma_{\beta\alpha}(\vec{r}_{a}-\vec{r}_{b})= k_{0}^{2}
\\
   \times \int \bra \grad_{\vec{r}_{a}}\beta(\vec{r}_{a})\cdot\grad_{\vec{r}_{a}}[\delta(\vec{r}_{a}-\vec{r}_{1})]
      G_{0}(\vec{r}_{1}-\vec{r}_{b})\alpha(\vec{r}_{b})\ket\ud^{3}{r}_{1}.
\end{multline}
Performing two integrations by parts and using the properties of the Dirac distribution yields:
\begin{equation} \label {Sigma_numu}
   \Sigma_{\beta\alpha}(\vec{r}_{a}-\vec{r}_{b})=
      k_{0}^{2}\grad_{\vec{r}_{a}}G_{0}(\vec{r}_{a}-\vec{r}_{b})\cdot\grad_{\vec{r}_{a}}C_{\beta\alpha}(\vec{r}_{a}-\vec{r}_{b}).
\end{equation}
Taking advantage of the assumed radial symmetry, we obtain:
\begin{equation} 
   \Sigma_{\beta\alpha}(x)=k_{0}^{2}\frac{\partial G_{0}}{\partial x} \frac{\partial C_{\alpha\beta}}{\partial x}.
\end{equation}

\subsection{Third term}\label{contrib_munu}
% ---------------------

Similarly to the second term, the third one implies both $\beta$ and $\alpha$:
\begin{multline}
   \Sigma_{\alpha\beta}(\vec{r}_{a}-\vec{r}_{b})= k_{0}^{2}
\\
   \times \int \bra
   \alpha(\vec{r}_{a})G_{0}(\vec{r}_{a}-\vec{r}_{2})\grad_{\vec{r}_{2}}[\beta(\vec{r}_{2})]
      \cdot\grad_{\vec{r}_{2}}[\delta(\vec{r}_{2}-\vec{r}_{b})]\ket\ud^{3}{r}_{2}.
\end{multline}
Again, performing two integrations by part and using the properties of the Dirac distribution, we obtain:
\begin{equation} \label {Sigma_munu}
   \Sigma_{\alpha\beta}(\vec{r}_{a}-\vec{r}_{b})=
      -k_{0}^{2}
      \grad_{\vec{r}_{b}}\cdot\left[G_{0}(\vec{r}_{a}-\vec{r}_{b})\grad_{\vec{r}_{b}}\left\{C_{\alpha\beta}(\vec{r}_{a}-\vec{r}_{b})\right\}\right].
\end{equation}
Since all functions involved here have radial symmetry, the expression above simplifies into
\begin{equation}
   \Sigma_{\alpha\beta}(x)=-\frac{k_{0}^{2}}{x^{2}}
      \frac{\partial}{\partial x} \left[x^{2}G_{0}(x) 
      \frac{\partial C_{\alpha\beta}} {\partial x}\right].
\end{equation}

\subsection{Fourth term}\label{contrib_nunu}
% ----------------------

The last term is the most complicated one:
\begin{multline}\label{Sigma_beta_beta}
   \Sigma_{\beta\beta}(\vec{r}_{a}-\vec{r}_{b})=
      \iint \bra \grad_{\vec{r}_{a}} \beta(\vec{r}_{a})\cdot\grad_{\vec{r}_{a}}[\delta(\vec{r}_{a}-\vec{r}_{1})]\right.
\\
   \left. \times G_{0}(\vec{r}_{1}-\vec{r}_{2})\grad_{\vec{r}_{2}}\beta(\vec{r}_{2})
      \cdot\grad_{\vec{r}_{2}}[\delta(\vec{r}_{2}-\vec{r}_{b})]\ket \ud^{3}{r}_{1}\ud^{3}{r}_{2}.
\end{multline}
\Eq{Sigma_beta_beta} contains a product of two scalar products, which can be written as a tensorial product. Again, two
integrations by parts and the properties of the Dirac distribution yield:
\begin{multline} \label {Sigma_nunu}
   \Sigma_{\beta\beta}(\vec{r}_{a}-\vec{r}_{b})=
\\
      -\grad_{\vec{r}_{b}}\cdot \left[\grad_{\vec{r}_{b}}\otimes\grad_{\vec{r}_{a}}
      \left\{C_{\beta\beta}(\vec{r}_{a}-\vec{r}_{b})\right\}
      \grad_{\vec{r}_{a}}G_{0}(\vec{r}_{a}-\vec{r}_{b})\right].
\end{multline}
Given the radial symmetry, in three dimensions we have:
\begin{equation}
   \Sigma_{\beta\beta}(x)=
      -\grad\cdot\left[\grad\otimes\grad\left\{C_{\beta\beta}(x)\right\}\grad G_{0}(x)\right].
\end{equation}
The tensorial product between the two gradients is a Hessian matrix. In spherical coordinates and for a function with
radial symmetry, we have~\cite{MASI-2007}:
\begin{equation}
   \grad\otimes\grad C_{\beta\beta}=
   \begin{bmatrix}
      \dfrac{\partial^{2}C_{\beta\beta}}{\partial x^{2}} & 0 & 0
   \\
      0 & \displaystyle\dfrac{1}{x} \dfrac{\strut\partial C_{\beta\beta}}{\strut\partial x} & 0
   \\
      0 & 0 & \displaystyle\dfrac{1}{x} \dfrac{\strut\partial C_{\beta\beta}}{\strut\partial x}
   \end{bmatrix}.
\end{equation}
Hence:
\begin{equation} 
   \Sigma_{\beta\beta}(x)=-\frac{1}{x^{2}}\frac{\partial}{\partial x}\left[x^{2}\frac{\partial^{2} C_{\beta\beta}}{\partial
      x^{2}}\frac{\partial G_{0}}{\partial x}\right].
\end{equation}

\section{Results and discussion}
%==================================

In real 3-D space, the four contributions to the self-energy add up to give:
\begin{multline} \label {Sigma_total_x}
   \Sigma(x)= k_{0}^{4}\sigma_{\alpha}^{2}G_{0}c_{\alpha \alpha}
      -k_{0}^{2}\sigma_{\alpha}\sigma_{\beta}G_{0}[c^{\prime\prime}_{\alpha \beta}+2{c^{\prime}_{\alpha\beta}}/x] 
  \\
      -\sigma_{\beta}^{2}[c^{\prime\prime}_{\beta\beta}G_{0}^{\prime\prime}
         +c^{\prime\prime\prime}_{\beta \beta}G_{0}^{\prime}+2c^{\prime\prime}_{\beta \beta}G_{0}^{\prime}/x].
\end{multline}

For simplicity the $x$-dependence of $G_{0}$ and of the correlation coefficients have been omitted, and the prime means
derivation with respect to $x$.

If $\sigma_{\beta}=0$, the self energy \eq{Sigma_total_x} is reduced to 
the usual (i.e., scalar only) term $\Sigma_{\alpha \alpha}$. In the general case where heterogeneity is such that a scalar ($\alpha$) and an operator ($\beta$) term coexist, it is not obvious to determine the orders of magnitude of the additional terms in the self-energy, since they involve five physical parameters: two variances and three correlation lengths. In order to highlight the importance of the additional terms relatively to the first one, we focus on the most simple case where $\sigma_{\alpha}=\sigma_{\beta}=\sigma$ and $c_{\alpha\alpha}=c_{\alpha\beta}=c_{\beta\beta}$. The variance $\sigma^{2}$ appears as a mere multiplicative term, and from a
physical point of view everything will depend on the typical correlation length $\ell_c$. Exponential and gaussian correlation functions were tested, and similar trends were obtained. We only give the result for the exponential case in 3-D, which entails simpler analytical expressions. The 2-D case is dealt with in the Appendix (section \ref{Appendix:2D}).

In the case of an exponentially-correlated disorder, we have
\begin{equation}
   c_{\alpha\alpha}(x)=\exp\left[-\frac{x}{\ell_{c}}\right].
\label{eq:alpha}
\end{equation}
In 3-D, the calculation of $\tilde{\Sigma}$ yields:
\begin{align}
   \tilde{\Sigma}(k_{0})=-\frac{\sigma^2}{\ell_c^2}
   &\left[\frac{(\uu)^4+j\uu}{1-2j\uu} \right.\nonumber
   \\
   \label{Sigma(k0)Expo}
   &-\left.\frac{1-2(\uu)^2}{\uu}\arctan\left(\frac{\uu}{1-j\uu}\right)\right]   
\end{align}

As long as $\uu\gg1$ (\ie the correlation length is much larger than the wavelength, the first term dominates:
\begin{equation}
   \tilde{\Sigma}(k_{0}) \sim \tilde{\Sigma}_{\alpha\alpha}(k_0)
   =-\frac{\sigma^{2}}{\ell_c^{2}}\frac{(\uu)^{4}}{1-2j\uu}.
\end{equation}

Therefore at high frequencies, even though the scalar and operator parts have equal importance in the random potential ($\sigma_\alpha=\sigma_\beta$), considering the usual wave equation with a space-dependent wave speed $c(\vec{r})$ instead of $c_0$ is legitimate to determine the coherent pressure field. However it becomes completely wrong as soon as $\uu$ is comparable to unity. In that case, the impact of the three additional terms ($\tilde{\Sigma}_{\alpha\beta}$, $\tilde{\Sigma}_{\beta\alpha}$ and $\tilde{\Sigma}_{\beta\beta}$) on the effective wavenumber $\keff$ and particularly the scattering mean-free path $\ell_{s}=1/[2\imag(\keff)]$ can be far from negligible.
More precisely, the difference is less than $6\,\%$ for $\uu>10$; but below $\uu\sim1.5$, the three additional terms in the self-energy are larger than the first one. As a result, at low frequencies the actual mean-free path can be nearly five times smaller than expected!  The exact ratio is $13/3$; the same behavior was obtained in the case of a gaussian-correlated disorder. Interestingly, it can be shown that the $13/3$ ratio is independent of the correlation function (as long as its second-order moment is finite, see Appendix, paragraph \ref{Appendix:3D}).
 
\begin{figure}[!htbf]
   \centering
   \psfrag{x}[c]{$k_0\ell_{c}$}
   \psfrag{y}[c]{$\sigma^2 k_0\ell_s$}
   \includegraphics[width=\linewidth]{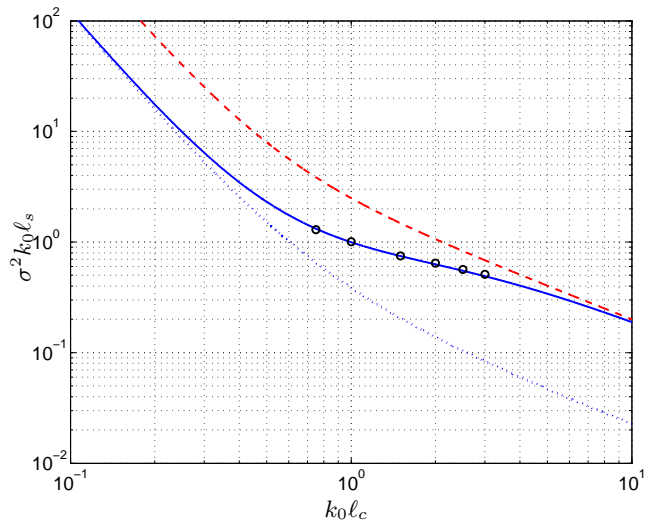}
   \caption{\label{Leg1} (Color online) Scattering mean-free path as a function of $k_0\ell_{c}$, with (solid line) and without (dashed line) the additional terms in the self-energy, for an exponentially-correlated disorder. Results from the numerical experiments are indicated by black circles. For comparison, the fluid limit of the model from Ref.\onlinecite{TURNER-2001} is also plotted (dotted line).}
\end{figure}

As an illustration, \figs{Leg1}{Leg2} compare the scattering mean-free paths obtained with ($\ell_s$) and without ($\ell_s^{(\alpha\alpha)}$) the additional terms.

Note that care should be taken when taking the low-frequency limit; the on-shell approximation usually requires ${\tilde{\Sigma}(k_{0})}$ to be much smaller than $k_{0}^{2}$, hence (from \eq{Sigma(k0)Expo}) when $\uu\rightarrow0$ the results are consistent only if the variance is kept such that $\sigma^{2}\ll (\uu)^{2}$. Interestingly, in the standard (scalar) case at low frequency the same condition implies that $\sigma^{2}\ll {1}/{(\uu)^{2}}$. This means that whatever the fluctuations $\sigma$, a weak disorder approximation (${\tilde{\Sigma}(k_{0})} \ll k_{0}^{2}$) is always fulfilled at zero frequency if the random operator is purely scalar. This is no longer true when the operator term cannot be neglected: for a finite $\sigma$, there is a cut-off frequency (typically $\uu\sim\sigma$) below which ${\tilde{\Sigma}(k_{0})}$ is not small compared to $k_{0}^{2}$. 

In order to test the validity of the theoretical results above, we have performed numerical simulations of the inhomogeneous wave equations (\ref{Eq_base1} and \ref{Eq_base2}), using a finite-difference software developed in our lab~\cite{BOSSY-2004},\footnote{All information regarding the software are available on the website www.simsonic.fr}. Simulations were carried out for conditions typical of ultrasonic experiments. The reference (unperturbed) medium was water ($c_0=1500\,$m/s $\rho_0=1000\,$kg/m$^3$) and the incoming waveform was a short pulse with a central frequency ranging from $1$ to $2\,$MHz. Using a random number generator, exponentially-correlated 3-D maps with $0$ mean and standard deviation $\sigma$ were fabricated and used for both $\alpha$ and $\beta$. For a given realization of disorder, once $\alpha(\vec{r})$ and $\beta(\vec{r})$ are determined the corresponding density, compressibility and sound speed are accessible from \eqs{def_alpha}{def_beta}. The typical correlation length was $\ell_c=0.240$\ mm, so that $k_0\ell_c=1$ at $1\,$MHz. Given the frequency spectrum, $k_0\ell_c$ spans typically between 0.75 and 3 in the numerical experiments. Further details on the simulations are given in the Appendix, section \ref{Appendix:Simul}.

\begin{figure}[!htbf]
   \centering
   \psfrag{x}[c]{$k_0\ell_{c}$}
   \psfrag{y}[c]{$\ell_s/\ell_s^{(\alpha\alpha)}$}   
   \includegraphics[width=\linewidth]{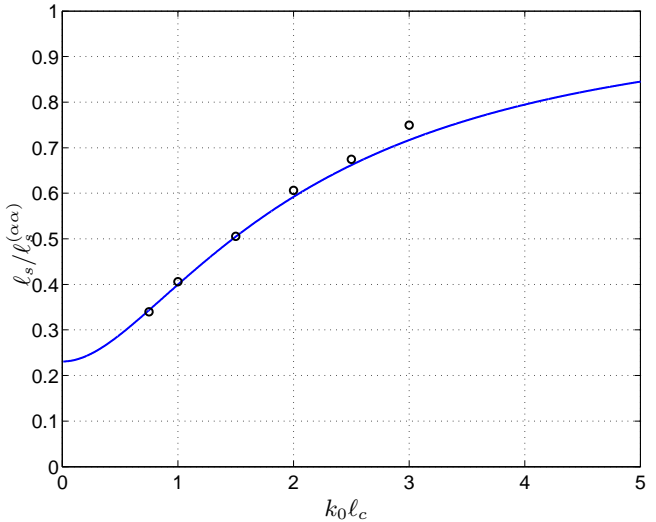}
   \caption{\label{Leg2} (Color online) Ratio of the scattering mean-free paths obtained with and without the additional terms in the self-energy as a function of $k_0\ell_{c}$, for an exponentially-correlated disorder. Results from the numerical experiments are indicated by black circles. The value at zero frequency is 13/3.}
\end{figure}

A plane wave is launched from on side of the random medium ($z=0$) in the $z$ direction. The resulting pressure is measured at every grid point $(x,y,z)$ and time $t$. A robust estimation of the coherent wave is obtained by a two-step average. For each realization of disorder, the pressure field is averaged in the $(x,y)$ plane, as would do a plane detector perpendicular to the initial direction of propagation. Secondly, ensemble-averaging  is performed over at least $25$ different realizations of disorder. As a result we obtain an estimate of the coherent pressure field $\bra p(z,t) \ket$ as a function of depth $z$ and time $t$. In all the numerical experiments, the total thickness of the map was at least $3\ell_s$ and we always ensured that the measured wavefront was an accurate estimator of the coherent field (that is to say remaining random fluctuations could be considered as negligible). A digital Fourier transform is performed; the coherent field's intensity $I_c=|\bra P(z,\omega) \ket|^2$ is found to decay exponentially with $z$. An estimation of the scattering mean-free path is obtained by a linear fit of $\log(I_c)$ with $z$, at each frequency. Simulating both types of media separately (either $\sigma_\beta=0$ hence no operator term in the random potential, or $\sigma_\beta=\sigma_\alpha=\sigma$), we plotted in Fig.~\ref{Leg2} the ratio of the corresponding mean-free paths. The results are in very good agreement with the analytical results presented earlier and support the validity of the theoretical analysis.
The numerical results were also compared to an other model derived from acoustics in polycrystals with randomly varying
elastic properties, but macroscopically isotropic~\cite{TURNER-2001}. In the limit where the second Lam\'{e} coefficient $\mu$ tends to 0 (no shear stress), the results should be valid for the case of an inhomogeneous liquid. Interestingly, this model does predict the 13/3 factor at 0 frequency, yet it yieds incorrect results at higher frequencies, especially above $k_0\ell_c>0.1$. The essential reason is that in the solid model, the fluctuations in mass density and elastic constants are assumed to be very weak from the very beginning (i.e., the linearized equations of elastodynamics). In our approach, fluctuations are not necessary weak initially, what is considered as weak is the second-order terms in the developement of the self-energy (Bourret approximation). The weak fluctuation limit, if necessary, is only taken afterwards. Assuming that the fluctuations are weak from the beginning amounts to misestimate some of the additional terms in the self energy. In the case of a fluid with $\sigma_\alpha=\sigma_\beta$, our results indicate that they cannot be discarded, no matter how weak fluctuations are; hence the results presented here for an inhomogeneous fluid cannot be seen as a particular case of the solid model. Note that we do not claim at all that the solid model in Ref.~\onlinecite{TURNER-2001} is wrong: it is very well suited for polycrystals (e.g., coarse-grain steel), in which fluctuations of mass density and elasticity are indeed very weak compared to there mean value.

\begin{figure}[!htbf]
   \centering
   \psfrag{x}[c]{$k_0\ell_{c}$}
   \psfrag{y}[c]{$n$}   
   \includegraphics[width=\linewidth]{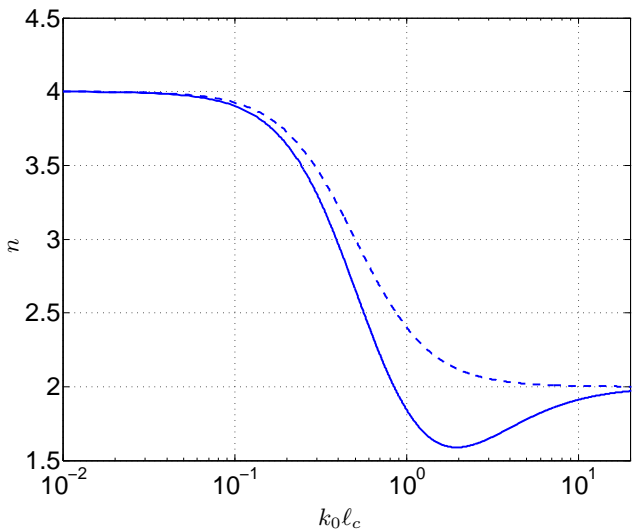}
   \caption{\label{Leg3} (Color online) Characteristic exponent $n$ obtained with (solid line) and without (dashed line) the additional terms in the self-energy as a function of dimensionless frequency $k_0\ell_{c}$, for an exponentially-correlated disorder.}
\end{figure}

It is also interesting to plot the exponent $n=-\omega/\ell_s\,\ud\omega/\ud\ell_s$ as a function of frequency (Fig. \ref{Leg3}). Indeed, since a power-law dependence of the attenuation length is often assumed, $n$ commonly serves as an indicator of the scattering regime. In both cases, $1/\ell_s$ is found to be proportional to $\omega^4$ at low frequency and $\omega^2$ at high frequency. These two trends are usually referred to the Rayleigh and stochastic regimes and are used to characterize scattering media based on the measured dependence of acoustic attenuation with frequency. Fig. \ref{Leg3} shows how misleading the omission the additional terms in the wave equation can be, especially at intermediate frequencies ($k_0\ell_c\sim1$): the exponent can be 35\% lower than expected.
  
However, as to the velocity of the coherent field, the effect of the additional terms is very limited since the real part of the wave vector is
\begin{equation}
   \real(\keff)\approx k_{0}-\frac{\real\left[{\tilde{\Sigma}(k_{0})}\right]}{2k_{0}},
\end{equation}
which will always remain close to $k_{0}$ within the on-shell approximation.

\section{Conclusion}
% ==================
Starting from the wave equation for the acoustic pressure in an heterogeneous and non-dissipative fluid, we have calculated the coherent wave, taking into account spatial variations of both density and compressibility such that the relevant random potential contains both a scalar and an operator part, $\alpha$ and $\beta$. The calculation is based on the diagrammatic approach of multiple scattering, within Bourret and on-shell approximations. Interestingly, the results show that discarding the random operator term (as is usually done when treating the problem as Helmholtz' equation with a space-dependent wavespeed $c(\vec{r})$ amounts to overestimate the scattering mean-free path by up to a factor of five when the fluctuation of $\alpha$ and $\beta$ have similar magnitude. The error is particularly large at low frequencies, when the correlation length is comparable to or smaller than the wavelength. The theoretical analysis has been conducted in two and three dimensions, and validated by numerical experiments. Though the results presented here are theoretical and rather academic, we believe they are of importance for all practical applications involving multiple scattering of acoustic waves e.g., characterization inhomogeneous media. Moreover, from a theoretical point of view, the scattering mean-free path is the basic ingredient to describe universal wave phenomena in complex media, such as coherent backscattering, ballistic-to-diffuse transition, radiative transport of energy etc. It is therefore crucial to determine it properly.

\begin{acknowledgements}
   This work was supported by the \emph{Agence Nationale de la Recherche} (ANR-11-BS09-007-01, Research Project
   DiAMAN), LABEX WIFI (Laboratory of Excellence ANR-10-LABX-24) within the French Program ``Investments for the
   Future'' under reference ANR-10-IDEX-0001-02 PSL$^{\ast}$ and by \'Electricit\'e de France R\&D.
\end{acknowledgements}

\section{Appendix}
% ==================

\subsection{3-D calculations}\label{Appendix:3D}
%---------------------------
Assuming that the correlation functions $C_{\alpha\alpha}$, $C_{\alpha\beta}$ and $C_{\beta\beta}$ are identical, we have 
\begin{multline} \label {Sigma_total_2}
   \Sigma(x)=k_0^4G_{0}C
      -k_0^2G_0[C^{\prime\prime}+2C^{\prime}/x]
      \\
      -[C^{\prime\prime}G_0^{\prime\prime}
      +C^{\prime\prime\prime}G_0^{\prime}+2C^{\prime\prime}G_0^{\prime}/x]
      \\
      =G_0[k_0^4C-2k_0^2C^{\prime}/x+(1/x-jk_0)C^{\prime\prime\prime}].
\end{multline}

Given the radial symmetry, the 3-D Fourier transform of $\Sigma$ is
\begin{equation} 
   \tilde{\Sigma}(k_0)=\frac{4\pi}{k_0} \int\limits_{0}^{\infty} \Sigma(x)\sin(k_0x)x\ud x.
\end{equation}

Hence the imaginary part:
\begin{multline} \label{BeforeDL}
   \imag \tilde{\Sigma}(k_0)=\frac{1}{k_0}
   \int\limits_{0}^{\infty}
        \left[k_0^4C -2k_0^2C^{\prime}/x+C^{\prime\prime\prime}/x\right]\sin^2(k_0x)\ud x
         \\
         - \frac{1}{2}\int\limits_{0}^{\infty}C^{\prime\prime\prime}\sin(2k_0x)\ud x.
\end{multline}

In order to study its behavior in the low-frequency regime ($k_0x\rightarrow0$) a Taylor expansion of the sines up to the sixth order followed by integrations by parts are performed. It yields
\begin{equation} \label{DL}
   \imag \tilde{\Sigma}(k_0)
   \rightarrow  k_0^5
   \int\limits_{0}^{\infty}
        \frac{13}{3}x^2C(x)\ud x.
\end{equation}

If the additional terms due to the random operator are neglected, \Eq{BeforeDL} reduces to 
\begin{align} 
   \imag \tilde{\Sigma}(k_0)&=\frac{1}{k_0}
   \int\limits_{0}^{\infty}k_0^4C \sin^2(k_0x)\ud x
   \\
   &\rightarrow  k_0^5\int\limits_{0}^{\infty}x^2C(x)\ud x.
\end{align}

As a consequence, in the low frequency limit $\uu\rightarrow0$, the ratio of the mean-free path calculated with
($\ell_s$) or without ($\ell_s^{(\alpha\alpha)}$) the additional terms is $13/3$. This ratio does not depend on the
precise shape of the correlation function $C(x)$, as long as its second-order moment is finite. 

The final results given and plotted in the paper were established for an exponentially-correlated disorder. In the gaussian case where $C(x)=\sigma^2\exp(-x^2/\ell_{c}^2)$, we obtain
\begin{multline} 
	\frac{\tilde{\Sigma}(k_0)}{k_0^2}=\sqrt{\pi}\frac{\sigma^2}{4}
	\left[
	jk_0\ell_c(9E-1)+\frac{8}{\sqrt{\pi}} \right.\\
	\left. +\frac{1}{k_0\ell_c}4j(3E-1)+\frac{1}{(k_0\ell_c)^2}\frac{8}{\sqrt{\pi}}
	+\frac{8j(E-1)}{(k_0\ell_c)^3} \right]. 
\end{multline}

In the expression above, we have introduced a dimensionless constant $E$:
\begin{equation} 
E=(1+\erf(jk_0\ell_c))e^{-k_{0}^2\ell_c^2}.
\end{equation} 

If the additional terms are neglected, we have
\begin{equation} 
    \frac{\tilde{\Sigma}(k_{0})}{k_{0}^2}=\sqrt{\pi}\frac{\sigma^2}{4}
	\left[ jk_0\ell_c(E-1)
	\right].
\end{equation} 

For the sake of simplicity, the ratio ($\ell_s/\ell_s^{(\alpha\alpha)}$) has not been plotted in the Gaussian case, but its general trend is very similar to the exponential case.

\subsection{2-D calculations}\label{Appendix:2D}
%---------------------------

In 2-D space, we have:  
\begin{eqnarray}
G_0(\vec{r}-\vec{r}_{s}) &=& \frac{-i}{4}\operatorname{H}_0^{(1)}(k_0|\vec{r}-\vec{r}_{s}|), \label{eq:Green2D}\\
\langle G(\vec{r}-\vec{r}_{s})\rangle &=& \frac{-i}{4}\operatorname{H}_0^{(1)}(\keff |\vec{r}-\vec{r}_{s}|).
\end{eqnarray}
$\operatorname{H}_0^{(1)}(x)$ is the Hankel function of the first kind and of order $0$. 
We still have four contributions to the self energy (Eq.~(\ref{eq:Self})). Assuming circular symmetry, with $x=|\vec{r_a}-\vec{r_b}|$ we have:
\begin{eqnarray*}
	\Sigma_{\alpha\alpha}(x) &=& k_0^4G_0(x)C_{\alpha\alpha}(x)\\
	\Sigma_{\beta\alpha}(x) &=& k_0^2\vec{\nabla}G_0(x)\cdot \vec{\nabla}C_{\beta\alpha}(x)\\
	\Sigma_{\alpha\beta}(x) &=& - k_0^2\vec{\nabla}\cdot \left[G_0(x)\vec{\nabla}C_{\alpha\beta}(x)\right]\\
	\Sigma_{\beta\beta}(x) &=& - \vec{\nabla}\cdot \left[\vec{\nabla}\otimes \vec{\nabla}C_{\beta\beta}(x)\vec{\nabla} G_0(x)\right].
\end{eqnarray*} 

The difference between the 2-D and 3-D cases lie in the expressions of the gradient, divergence and Hessian of a function with circular (or spherical) symmetry. In particular, $\Sigma_{\beta\beta}$ requires the Hessian of a circularly symmetric function in polar coordinates:

\begin{equation}
\grad\otimes\grad C_{\beta\beta}=
\begin{bmatrix}
\dfrac{\partial^{2}C_{\beta\beta}}{\partial x^{2}} & 0 
\\
0 & \displaystyle\dfrac{1}{x} \dfrac{\strut\partial C_{\beta\beta}}{\strut\partial x} 
\end{bmatrix}.
\end{equation}

As a whole, in 2-D the expression of the self energy (equivalent of \eq{Sigma_total_x} in 3-D) reads:
\begin{align}
\Sigma(x) =& k_0^4G_0(x)\sigma_\alpha^2c_{\alpha\alpha}(x)  \nonumber\\&-k_0^2 G_0(x)\sigma_\alpha\sigma_\beta\left[c''_{\alpha\beta}(x) + \frac{1}{x}c'_{\alpha\beta}(x) \right] \nonumber\\ 
&-\sigma_\beta^2\left[G'_0(x)c'''_{\beta\beta}(x) + \left(G''_0(x)+\frac{1}{x}G'_0(x)\right)c''_{\beta\beta}(x)\right]\nonumber \\
&&
\end{align}

Using \eq{eq:Green2D} along with differentiation and recurrence properties for Bessel and Hankel functions (\cite{ABRAMOWITZ} page 361), it is straightforward to obtain:
\begin{eqnarray}
\Sigma(x) &=& G_0(x) \bigg[ k_0^4\sigma_\alpha^2c_{\alpha\alpha}(x) \nonumber\\
 &&\left. - k_0^2\sigma_\beta\left(\sigma_\alpha c''_{\alpha\beta}(x) + \frac{1}{x}\sigma_\alpha c'_{\alpha\beta}(x)- \sigma_\beta c''_{\beta\beta}(x) \right) \right] \nonumber\\
&& - G_0'(x)\sigma_\beta^2c'''_{\beta\beta}(x)
\label{eq:SelfN}
\end{eqnarray}

And for identical correlation functions $C_{\alpha\alpha}$, $C_{\alpha\beta}$ and $C_{\beta\beta}$, we have:
\begin{equation}
\Sigma(x)=\sigma^2G_0(x)\left[k_0^4c(x)-\frac{k_0^2}{x}c'(x)\right]-\sigma^2G_0'(x)c'''(x).
\end{equation}

Once an analytical expression for $\Sigma(x)$ is obtained, we have to calculate its spatial Fourier Transform in order to determine the effective wave number. In 2-D, the Fourier transform of a circularly symmetric function is the zero-order Hankel transform:
\begin{equation}
\tilde{\Sigma}(k)=2\pi\int_0^\infty \Sigma(x)\operatorname{J}_0(kx)x \dd x
\end{equation}
where $\operatorname{J}_0(x)$ is the cylindrical Bessel function of order $0$. Calculating the mean free path ($\ell_s=1/[2 \imag(\keff)]$) amounts to numerically evaluating three integrals:
\begin{eqnarray}
\tilde{\Sigma}(k) &= &\frac{-i\pi }{2}\left(k_0^4I_1 -k_0^2I_2+  k_0I_3\right)  \quad \mbox{where:}\label{eq:Self_op}\\
I_1 & = &\int_0^\infty C(x)H_0^{(1)}(k_0x)J_0(kx)x\dd x\\
I_2 & = & \int_0^\infty C'(x)H_0^{(1)}(k_0x)J_0(kx)\dd x\\
I_3 & = &  \int_0^\infty C'''(x)H_1^{(1)}(k_0x)J_0(kx)x\dd x
\label{SelfKK0}
\label{BeforeDL2}
\end{eqnarray}
whatever the shape of the correlation function $C(x)$.

In the low-frequency regime ($k_0x\rightarrow0$), a Taylor expansion of the Bessel functions followed by integrations by parts yield
\begin{equation} \label{DL}
   \imag \tilde{\Sigma}(k_0)
   \rightarrow -\frac{\pi}{2} k_0^4
   \int\limits_{0}^{\infty}
        \frac{9}{2}xC(x)\ud x.
\end{equation}

If the additional terms due to the random operator are neglected (\ie $I_2=I_3=0$), \Eq{BeforeDL2} reduces to 
\begin{align} 
   \imag \tilde{\Sigma}(k_0)&=\frac{4\pi}{k_0}
   \int\limits_{0}^{\infty}k_0^4C \sin^2(k_0x)\ud x
   \\
   &\rightarrow -\frac{\pi}{2} k_0^4\int\limits_{0}^{\infty}xC(x)\ud x.
\end{align}

Hence, in the low frequency limit $\uu\rightarrow0$, the ratio of the 2-D mean-free paths calculated with
($\ell_s$) or without ($\ell_s^{(\alpha\alpha)}$) the additional terms is $9/2$, as opposed to $13/3$ in 3-D. And again, this ratio does not depend on the precise shape of the correlation function $C(x)$, as long as its second-order moment is finite. 

%----------------------------
\subsection{Numerical simulations}\label{Appendix:Simul}
%---------------------------

\begin{figure}[!htbf]
	\centering
	%\psfrag{x}[c]{$k_0\ell_{c}$}
	%\psfrag{y}[c]{$\sigma^2 k_0\ell_s$}
	\includegraphics[width=0.6\linewidth]{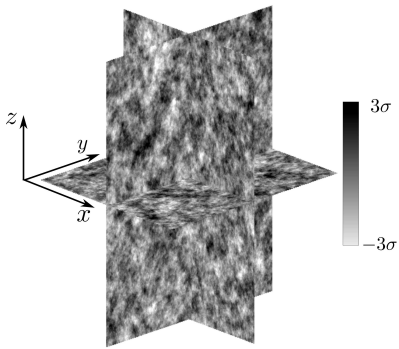}
	\caption{Example of an exponentially correlated disorder with gaussian statistics (zero mean, variance $\sigma^2$) in 3-D space.}
	\label{fig:SI_1}
\end{figure}

The acoustic wave propagation in heterogeneous  media is numerically simulated with Simsonic, a 3-D Cartesian FDTD approach to solve the elastodynamic equations. 3-D maps of the local wavespeed and mass density can be designed by the user. These maps define the propagation media at each grid point. By properly filtering a 3-D white noise of $N_x \times N_y \times N_z $ points with Gaussian statistics, it is possible to build a 3-D map exhibiting an exponentially correlated disorder:

$$C_{\alpha \alpha}(r)=\sigma^2\exp\left[-\frac{r}{\ell_c}\right]$$
where $r=\sqrt{x^2 +y^2+z^2}$ is the radial coordinate, $\ell_c$ the correlation length and $\sigma^2$ the variance. An example of one realization of the media is given in Fig.~\ref{fig:SI_1}. Values below $-3\sigma$ or above $3\sigma$ are truncated. Various uncorrelated  realizations of disorder can be obtained by repeating the procedure.
The same map has been employed for $\alpha$ and $\beta$, so that $\alpha(\vec{r})=\beta(\vec{r})$ for each realization of disorder. In that case, we have $C_{\alpha\alpha}=C_{\beta\beta}=C_{\alpha\beta}$ which corresponds to the theoretical example detailed in the paper. Using independent, or partially correlated maps, or with differents variances or correlation lengths for $\alpha$ and $\beta$ could also be possible to investigate all possibilities.

The correlation length was set at $\ell_c=0.240\,$mm, so that $k_0\ell_c=1$ for a driving frequency $f=1 \,$MHz. The variance $\sigma^2$ ranged between 1$\%$ and 4$\%$. Various simulations were carried out in order to calculate the scattering mean free paths for frequencies in the range $k_0\ell_c \in [0.75 \,,\, 3]$ by changing the central frequency of the incoming pulse between 1 and 2 MHz. In order to avoid an additional numerical dissipation of the acoustic energy, it is important to resolve both the correlation length $\ell_c$ and the wavelength $\lambda$ with at least 10 grid points in all directions. Furthermore the CFL (Courant-Friedrichs-Lewy) condition is to be respected based on the maximum propagation speed $c_\text{max}=c_0/\sqrt{1-3\sigma}$ in the medium. Perfectly matched layers (PML) were implemented outside the scattering region to ensure absorbing boundary conditions. Typically more than 80 Go of RAM were required and a multi-threaded parallel version of Simsonic3D (OpenMP) was needed to perform these large-scale simulations. 

\begin{figure}[!htbf]
	\centering
	%\psfrag{x}[c]{$k_0\ell_{c}$}
	%\psfrag{y}[c]{$\sigma^2 k_0\ell_s$}
	\includegraphics[width=0.6\linewidth]{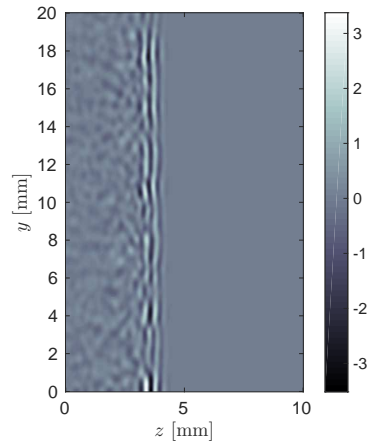}
	\caption{A pulsed plane wave is generated at $z=0\,$mm and propagates along the $z$-axis. The snapshot is taken at $t=4\,\mu$s, in the $x=0$ plane. The resulting pressure is in arbitrary units.}
	\label{fig:SI_2}
\end{figure}

As a typical example, a snapshot of the propagating wavefront in the $(y\,,z)$ plane is given in \fig{fig:SI_2}. 

Prior to an ensemble average of the acoustic pressure field over $N$ realizations of disorder, $p$ is first spatially averaged along the $(x,y)$ plane (under the hypothesis of spatial ergodicity) to obtain  $\bar{p}_i(z,t)$. Then the final mean field estimator reads:
\begin{equation}
\langle p (z,t)\rangle = \frac{1}{N}\sum_{i=1}^N \bar{p}_i(z,t) 
\label{eq:estimator}
\end{equation}
In all simulations we ensured $N$ to be large enough in order that \eq{eq:estimator} represents an accurate estimator of the 
mean coherent pressure field and that remaining random fluctuations can be neglected. 

\begin{figure}[!htbf]
	\centering
	%\psfrag{x}[c]{$k_0\ell_{c}$}
	%\psfrag{y}[c]{$\sigma^2 k_0\ell_s$}
	\includegraphics[width=0.85\linewidth]{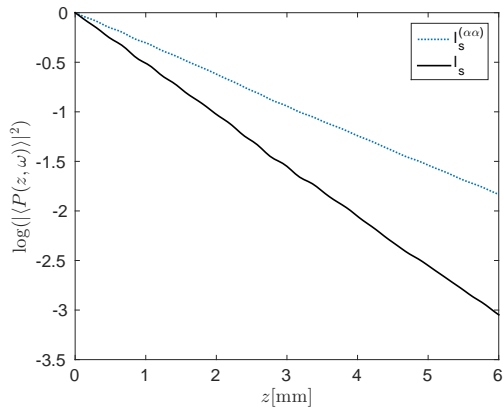}%{FigLogCoherent.eps}
	\caption{(Color online) Decay of $\log|\langle P (z,\omega)\rangle|^2$ versus distance $z$, at $f=2\,$MHz and $\sigma=0.2$. A linear fit of this data gives an estimation of the scattering mean free path in the scalar case (dashed line) and in the operator case (continuous line).}
	\label{fig:SI_3}
\end{figure}

From the mean pressure field we can calculate the acoustic intensity in the frequency domain, $|\langle P (z,\omega)\rangle|^2$. As seen in \fig{fig:SI_3}, the acoustic intensity decays exponentially; a linear fit of its logarithm gives an estimation of the mean free paths ($\ell_s$ or $\ell_s^{(\alpha\alpha)}$) at a given frequency. In all simulations, we ensured that the propagation distance was at least three times larger than the scattering mean-free path, so that the decay of intensity is significant.

%----------------------------
\subsection{Fluid limit of the solid model}
%---------------------------

In Ref.~\onlinecite{TURNER-2001}, the self energy is expressed in terms of fluctuations of mass density $\rho$ and Lam\'{e} coefficients $\lambda$ and $\mu$. The liquid limit is taken by setting $\mu=0$ and $\chi=1/\lambda$. Since the fluctuations of all parameters relative to their mean are assumed to be very small, \eqs{def_alpha}{def_beta} can be differentiated to obtain linear relations  between the two pairs of variables. This leads to $\sigma_\beta=\sigma_\rho$ and $\sigma_\lambda=\bra(\alpha+\beta)^2\ket$. In the scalar case, $\sigma_\beta=0$, then $\sigma_\rho=0$ and $\sigma_\alpha=\sigma_\lambda$. If the operator term is taken into account and and $\alpha(\vec{r})=\beta(\vec{r})$ then $\sigma_\beta=\sigma_\rho=\sigma$ and $\sigma_\alpha=\sigma_\lambda/2=\sigma$. From Eq. (37) in Ref.~\onlinecite{TURNER-2001}, we infer
\begin{multline}
	\label{Turner_scal}
	\dfrac{1}{k_0\ell_s^{(\alpha\alpha)}}=\sigma^2(\uu)^3\int_{-1}^{1}\dfrac{\dd x}{[1+2(\uu)^2(1-x)]^2}
	\\=\sigma^2\dfrac{2(\uu)^3}{1+4(\uu)^2}
	%\nonumber
\end{multline}
in the scalar case, and
\begin{multline}
\label{Turner_op}
\dfrac{1}{k_0\ell_s}=\sigma^2(\uu)^3\int_{-1}^{1}\dfrac{(x+2)^2}{[1+2(\uu)^2(1-x)]^2}\dd x
\\
%=\sigma^2\dfrac{(\uu)^2+8(\uu)^4+18(\uu)^6-[\frac{1}{4}+\frac{5}{2}(\uu)^2+6(\uu)^4]\ln(1+4(\uu)^2}{(\uu)^3(1+4(\uu)^2)}
%\\
=\sigma^2\dfrac{1+8(\uu)^2+18(\uu)^4}{\uu(1+4(\uu)^2)}
\\
-\sigma^2\dfrac{[1+10(\uu)^2+24(\uu)^4]\log(1+4(\uu)^2)}{4(\uu)^3(1+4(\uu)^2)}
\end{multline}
in the operator case.
\Eq{Turner_scal} is exactly our result in the scalar case (see \eq{ls_cas_expo}). However, in the operator case, \eq{Turner_op} and \eq{Sigma(k0)Expo} disagree, except at zero frequency, as was discussed earlier and shown in \fig{Leg1}.

% Biblio

\end{document}